\documentclass[aps,prb,superscriptaddress,floatfix,twocolumn,showpacs]{revtex4-1}

\usepackage{amsmath,amssymb}
\usepackage{graphicx}
\usepackage{color}

\def\bea{\begin{eqnarray}}
\def\eea{\end{eqnarray}}
\def\ben{\begin{equation}}
\def\een{\end{equation}}
\def\benu{\begin{enumerate}}
\def\enu{\end{enumerate}}

\def\br{{\bf r}}

\def\bu{{\bf u}}

\begin{document}

\title{{Bonds, lone pairs, and shells probed by means of on-top dynamical correlations}}

\author{Stefano Pittalis}
\email[]{stefano.pittalis@nano.cnr.it}
\affiliation{CNR-Istituto Nanoscienze, Via Campi 213 A, I-41125 Modena, Italy}

\author{Daniele Varsano}
\affiliation{CNR-Istituto Nanoscienze, Via Campi 213 A, I-41125 Modena, Italy}

\author{Alain Delgado}
\affiliation{Department of Physics, University of Ottawa, Ottawa, Ontario K1N 6N5, Canada}

\author{Carlo Andrea Rozzi}
\affiliation{CNR-Istituto Nanoscienze, Via Campi 213 A, I-41125 Modena, Italy}

\begin{abstract}
The Electron Localization Function (ELF)  by Becke and Edgecombe [J. Chem. Phys. {\bf 92}, 5397 (1990)] is routinely adopted as a  descriptor of atomic shells and covalent bonds. 
Since the ELF and its related quantities find useful exploitation also in the  construction of modern density functionals,
the interest in complementing the ELF is linked to both the quests of improving  electronic structure descriptors and density functional approximations.
The ELF uses information which is available by considering parallel-spin electron pairs in single-reference many-body states. 
In this work, we complement this construction with information obtained by considering antiparallel-spin pairs 
whose  short-range correlations  are modeled by a density functional approximation. As a result, the approach
requires only a contained computational effort. 
Applications 
to a variety of systems show that, in this way, we  gain a spatial description of the bond in H$_2$ (which is not available with the ELF)
together with some trends not optimally captured by the ELF in other prototypical situations.  
\end{abstract}

\maketitle

\section{Introduction}\label{Sec1}


It is  a great pleasure to present a work within a special issue celebrating  
the 65th birthday of Hardy Gross. The characterization of bonds, lone pairs, and shells in atomistic systems is one of the topics to which Hardy has dedicated particular consideration in relation to density functionals.  
Strikingly, picturing bonds in molecules by means of  Lewis' structures, in terms of shared valence electrons, does not demand any knowledge of quantum mechanics, if a reference to relevant empirical facts is available \cite{Lewis}. 
The task becomes challenging when we look for quantitative pictures deduced from first principles \cite{Coulson55}.

Solutions of the Schr\"odinger equation can reproduce Lewis structures in terms of 
overlapping (possibly hybridized) atomic-like orbitals of Valence Bond Theory. But other solutions obtained within Molecular Orbital Theory
challenges the former description by introducing orbitals  which are delocalized  throughout the molecule.
A modern incarnation of Molecular Orbital Theory is the Kohn-Sham (KS) formulation of Density Functional Theory (DFT).
KS-DFT builds on the electron density. This quantity does not describe pairs but just a collection of {\em single} particles.
Yet, Bader showed that the Laplacian of the electron density can render the structure of
atomic shells and bonds \cite{Bader94}.  However, success was not systematic.
Following up other suggestions by Bader \cite{BaderXhole1,BaderXhole2}, Becke and Edgecombe 
proposed the Electron Localiation Fucntion (ELF). This is an indicator that exploits the antisymmetry of a many-electron state: specifically, 
if an electron is localized at a given point in space,  
another electron  cannot be brought at the same position with the  same spin \cite{BE90}.

The ELF
has become a valuable tool for visualizing the chemical structure of molecules and materials \cite{BE90,Savin92,Savin94}.
Its time-dependent extension
provides visual understanding of complex reactions involving the dynamics of excited electrons \cite{BMG}.
Analyses of electron localization are also important because they offer insights for improving  density functional constructions \cite{RG1}.
The ELF has been connected to the ability of density functional approximations to access local information on the spin-entanglement \cite{Ent},
to properties of the electronic stress tensor \cite{Tao}, and to time-dependent density functionals \cite{RG2}.
The interpretation of the ELF, however, is not free from difficulties \cite{SavinTheo05,SavinJCS05}.
In this work, we start with the well-known observation that the ELF does not provide spatially resolved information
for the Hydrogen molecule.
Alternative ``correlated'' ELFs have  already been suggested \cite{Kohout04,Kohout05,Silvi2010} but they usually  
entail quantities beyond KS-DFT and, thus, are computationally more demanding.
In this work, we  make  an
attempt to exploit only spin-restricted {\em occupied} KS [or Hartree-Fock (HF)] orbitals and quantities which are easily computable from them. 

The core of our idea is explained in full detail in Sec.~\ref{Scpl}. Essentially,
our proposal  is to detect spatial
inhomogeneous structures that can be related to covalent bonds,  atomic shells, and lone pairs 
in terms of the behavior of short-range correlations.
The procedure -- as long as {\em opposite spin}
dynamical correlations are concerned \cite{helgaker2000} -- is intended to probe the tendency of the electrons  to ``get together'' by  de-correlating.
More specifically, the procedure we propose rests on a deconstruction of a DFT model for dynamical correlations \cite{cB88}.
We show that within a contained computational effort, not only a spatial description of the bond in H$_2$ can be recovered but also other trends not optimally captured by the ELF can be 
 usefully complemented in other prototypical situations.

The manuscript is organized as follows: In Sec.~\ref{Sec2}, we recall some basic theoretical aspects and, after a brief introduction to the ELF, 
we describe the Coalescent (electron) Pair Locator (CPL) and tackle the problematic case of H$_2$. In Sec.~\ref{Sec3}, 
we test our procedure for atoms, small organic molecules, and the uniform electron gas.
In Sec.~\ref{Sec4}, we summarize the main points, draw the conclusions, and give outlooks.

\section{From the ELF  to a new indicator based on on-top dynamical correlations}\label{Sec2}

Let us start by considering the definition of the distribution of electron pairs in terms of the 
joint probability to find an electron at position  $\br_1$ with spin $\sigma_1$ and a second electron at position  $\br_2$ with spin $\sigma_2$:
\ben\label{p2}
P^{\sigma_1 \sigma_2}(\br_1,\br_2) = N(N-1) \int d3 d4... dN~ | \Psi(1,2,3,...,N)|^2\;.
\een 
Here, $\Psi(1,2,3,...,N)$ is a normalized wavefunction of $N$ interacting electrons and $1,2, ..., N$ stand for the combined spatial plus spin variables $(\br_1,\sigma_1), (\br_2,\sigma_2), ... , (\br_N,\sigma_N)$. 
A direct calculation of $P^{\sigma_1 \sigma_2}(\br_1,\br_2)$ requires knowledge of the wavefucntion, which, however, can  be  computed efficiently and accurately only for small systems.  To reduce the computational burden, in this work,
we regard $P^{\sigma_1 \sigma_2}(\br_1,\br_2)$ as a density functional and adopt a simple model for it.
 
We should also emphasize that, in order to picture bonds, one would like to visualize domains in which the probability of finding two electrons is maximum. It turns out that the pair density in Eq.~(\ref{p2}) is not the quantity that can be used alone. For a discussion of this point, we refer the reader to Refs. \cite{SavinJCC05,SavinCTC15}. 
Becke and Edgecombe  built the ELF by exploiting information on conditional probabilities \cite{BE90}. In our work, we use the pair density in combination with other simple quantities.

Next, we review the procedure of the original construction of the ELF. In this way,
quantities used throughout can be  introduced gradually and the complementariness of our approach should also become more apparent. Then, finally, we illustrate our procedure and results.
For convenience, we shall keep track of spin indices explicitly in our notation, even though
we consider closed-shell ground states and carry out spin-adapted calculations\footnote{Our Kohn-Sham (or Hartree-Fock) calculations are intended as means to compute  a ``descriptor'' of the electronic structure of the system under consideration and not necessarily to estimate the best possible total energies -- thus we avoid to spuriously break spin symmetry.}.

\subsection{ELF in a nutshell}

In a single Slater Determinant (SD), electron pairs  are correlated in same-spin configurations 
\ben\label{pss_1}
P^{\sigma \sigma}_{\rm SD}(\br_1,\br_2) =  \rho^{\rm SD}_\sigma(\br_1) \left[ \rho^{\rm SD}_\sigma(\br_2) + h^{\sigma}_{\rm x}(\br_1,\br_2)\right] \;
\een
but are uncorrelated in opposite-spin configurations
\ben\label{p2ssd}
P^{\alpha\beta}_{\rm SD}(\br_1,\br_2) =  \rho^{\rm SD}_\alpha(\br_1) \rho^{\rm SD}_{\beta}(\br_2)\;.
\een
In Eq.~(\ref{pss_1}), $h^{\sigma\sigma}_{\rm x}(\br_1,\br_2)$ is the exchange-hole (x-hole) function
\ben
h^{\sigma}_{\rm x}(\br_1,\br_2) = - \frac{ | \gamma_{\rm SD}^{\sigma}(\br_1,\br_2) |^2 }{ \rho_\sigma^{\rm SD}(\br_1) }\;,
\een
where 
\ben\label{rho1}
\gamma_{\rm SD}^{\sigma}(\br_1,\br_2) = \Sigma_{i} \psi_{i\sigma}^*(\br_1)\psi_{i\sigma}(\br_2)\;
\een
is the spin-dependent one-body reduced density matrix;
here, {the sum over states} is restricted to occupied $\sigma$-state single-particle orbitals $\psi_{i \sigma}$.
The particle density of a single SD is readily obtained from the spatial diagonal of $\gamma_{\rm SD}^{\sigma}$:
$\rho^{\rm SD}_\sigma(\br)=\gamma_{\rm SD}^{\sigma}( \br, \br)$. 
The x-hole is a consequence of the fact that electrons are Fermions: their state is antisymmetric under exchange of
particle coordinates, which  implies a non-vanishing correlation even when the Coulomb repulsions is (ideally) turned off.

Next, let us
consider the {\em conditional} (cond) probability of finding an electron at distance $u$ given that a first electron is located at $\br_1$ and assuming that both electrons have the same spin. This entails to divide the quantity defined in Eq.~(\ref{pss_1}) by $\rho^{\rm SD}_\sigma(\br_1)$ and set $\br_2 = \br_1 + \bu$. The information on the orientation of the pair can be disregarded by performing  a spherical average 
\footnote{Given a generic function $f({\br_1,\br_2})$, its spherical average at position $\br$ for a pair at distance $u$ may be defined as the integral 
$f (\br,u) := \frac{1}{4 \pi } \int d \Omega_{\bu}~ f ( \br , \br + \bu)$  
where $d \Omega_{\bu}$ is an element of the solid angle.} around the ``reference'' position $\br_1$.
Dropping the pedantic specification to a SD in our notation, for small $u$ one finds \cite{BE90}
\ben\label{key_b}
 P^{\sigma \sigma}_{\rm cond}  (\br, u) = \frac{1}{3} D_\sigma(\br) u^2 + \cdot\cdot\cdot\;,
\een
where
\ben
D_\sigma(\br) := \left[ \tau_\sigma(\br) - \frac{1}{4} \frac{ \left( \nabla \rho_\sigma(\br) \right)^2}{\rho_\sigma(\br) } \right]\;,
\een
and
\ben
 \tau_\sigma(\br)  := \Sigma_{i} | \nabla \psi_{i\sigma}(\br) |^2
\een
is the double of the (positively defined) spin-dependent kinetic energy density.
The expressions above may be evaluated using either HF or, ideally, exact KS single-particle orbitals. Finally, Becke and Edgecombe defined \cite{BE90}
\ben\label{ELF}
\mathrm{ELF}(\br): = \frac{1}{ 1 + \chi^2_\sigma(\br) }\;
\een 
where
\ben\label{chi}
\chi_\sigma(\br) = D_\sigma(\br) / D^{\rm unif}_\sigma(\br)\;,
\een
and
\ben
D^{\rm unif}_\sigma(\br) = \frac{3}{5}\left( 6 \pi^2 \right)^{2/3} \rho^{5/3}_\sigma(\br)
\een
is evaluated for a Fermi gas at the local spin density \footnote{The {\em exact} KS Slater determinant in (Spin-)DFT reproduces the interacting particle (spin-)density exactly.} $\rho_\sigma(\br)$.

\subsection{Coalescent Pair Locator}\label{Scpl}

The ELF nicely reproduces the shell structure of atoms and
well emphasizes covalent bonds in molecules \cite{BE90,Savinreview}.
However, at any internuclear distance, for H$_2$ the ELF is constant in space (${\rm ELF}=1$). This is so because only one orbital is occupied for each spin channel in H$_2$ (thus $D_\sigma(\br) = 0$).
Here, we show how a spatially resolved description of the bond in H$_2$ may be recovered. 
In the next section, we will show that the resulting procedure can also deal with other interesting prototypical cases.

Our idea is based on the following simple picture:
If the Coulomb repulsion is turned off, two electrons with opposite-spin can be placed on top of each other and may remain in that configuration. 
In reality, the effects of the electron-electron repulsion can increase dramatically as electrons approach each  other: the electrons correlate {\em dynamically} 
and the wavefunction develops a cusp at the electron coalescence \cite{helgaker2000}.
Although the separation between dynamical and static correlations is not universally defined, it is known to provide useful insights.
Coming to our main objective, we may consider a covalent bond as a region where on-top dynamical correlations can become locally relatively weak and thus
electron pairs can ``get together'' more favorably. Below, we empirically evaluate this picture.

\begin{figure}[t]
\begin{center}
\includegraphics[width=0.9 \linewidth,angle=0]{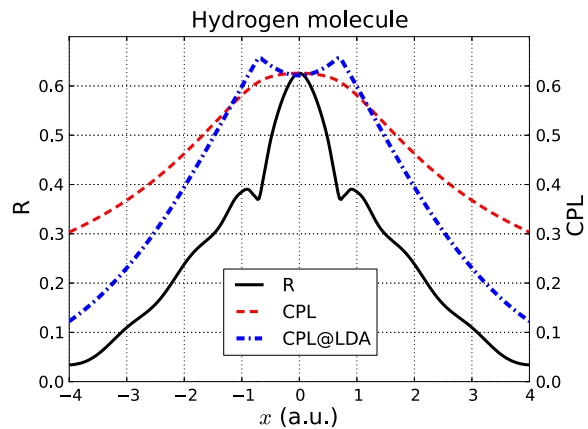}
\caption{The black-solid line shows the ratio $R$ [see Eq.~(\ref{R})] plotted for H$_2$ at an interatomic distance of 1.4 a.u., 
computed by using data from Ref. \cite{BPE98} (see in the main text). 
The dashed-red line shows the CPL  [see Eq.~(\ref{CPL})] with $c=0.292$. 
HF densities {calculated using Dunning} gaussian basis set aug-cc-pv5Z \cite{schuchardt2007basis} were employed to 
compute the Slater potential used in Eq.~(\ref{z}) and, thus, in Eq.~(\ref{CPL}).
The dot-dashed blue line refers to the CPL estimated according to Eq.~(\ref{uCPL}).
}\label{H2}
\end{center}
\end{figure}

The solid black curve in Fig.~\ref{H2} shows the ratio:
\ben\label{R}
\mathrm{R}(\br) = 
\frac{  P^{\alpha \beta}  (\br,u = 0) }{   P^{\alpha \beta}_{SD}  (\br,u = 0)}\;.
\een
for H$_2$ at its equilibrium distance. 
From Eq.~(\ref{p2ssd}), $P^{\alpha \beta}_{SD}  (\br,u = 0) = \rho^{\rm SD}_\alpha(\br) \rho^{\rm SD}_{\beta}(\br)$ and, we remind that,  ``SD'' 
stands for a spin-restricted Slater Determinant made of either HF or standard KS states.
In our calculation, we employed the data 
of the on-top exchange and exchange-correlation holes available from
Ref. \cite{BPE98} constructed from a Configuration-Interaction (CI) wavefunction calculation. 
Also, we have exploited the fact that the HF and the CI density are practically indistinguishable for the considered case
-- which, at the considered inter-distance of the nuclei, we have confirmed to be a valid approximation\footnote{Yet, it has been pointed out
in the literature that use of HF densities can produce counterintuitive features in the corresponding on-top hole \cite{BaerendsJCP10_a}.}.

The  salient point is: the solid-black curve of Fig.~\ref{H2} has a {\em maximum} where one would draw the shared ``dots''
in a Lewis diagram.
Instead, we remind that, $\mathrm{ELF} =1$ everywhere.

Equation (\ref{R}) involves the
{\em exact} on-top distribution of opposite-spin interacting pairs,  $P^{\alpha \beta}  (\br,u = 0)$.
It is useful to remind that the short-range behavior of this distribution is given by \cite{cB88}
\ben\label{SRud}
P^{\alpha \beta}  (\br,u) = A_{\alpha \beta}(\br)(1 + u) + \cdot\cdot\cdot \;,
\een
where $A_{\alpha \beta}$ depends on the reference point {\bf r}. 
As explained above, our aim is to focus on the contributions which are due to dynamical ({\em dyn}) correlations.
Thus, we replace the exact  $A_{\alpha \beta}(\br)$ with an approximation derived from a model
especially designed for dynamical  correlations \cite{cB88}. In this model, spin-resolved
correlation-hole (c-hole) functions are given for the electron-electron coupling interaction strengths $\lambda \in [0,1]$. For each $\lambda$, the c-hole functions contain zero electrons, while the corresponding x-hole
is $\lambda$ independent and contains {\em minus} one electron.
While the ultimate goal of the model is to derive an expression for the coupling averaged (DFT) c-hole, 
here we are interested in its implications at $\lambda = 1$. A straightforward calculation gives
\ben\label{pmodel}
A_{\alpha \beta}(\br)  =  \frac{\rho_\alpha(\br) \rho_\beta(\br)}{1 + z_{\alpha \beta}(\br)} \;,
\een
where $z_{\alpha \beta}(\br)$ is defined as
\ben\label{z}
z_{\alpha \beta}(\br) = c \left\{ | U_{{\rm x},{\alpha}}(\br) |^{-1} + | U_{{\rm x},{\beta}}(\br) |^{-1} \right\}\;
\een
with $c$ being a constant to be determined (see below) and
\ben\label{elc}
U_{{\rm x},{\sigma}}(\br)  =  \int d^3u~  \frac{ h^{\sigma}_{\rm x}(\br,\br + \bu) }{u}\;
\een
is the potential generated by the exchange-hole function -- also known as the Slater potential or 
a half of the spin-dependent exchange energy density in the standard gauge \cite{OEP}.
In passing, we note that $ U_{{\rm x},{\sigma}}(\br) = - 0.5 v_{\rm H}(\br)$ for systems with only two electrons in a singlet state and
the DFT exchange-only and the HF solutions are, in this case,  equivalent.

For any system for which we can run a DFT calculation, we can compute $U_{{\rm x},{\sigma}}$ and, thus, we can evaluate the following expression: 
\bea\label{CPL}
\mathrm{CPL}(\br)  & = & \frac{P^{\alpha \beta}_{dyn}  (\br,u = 0)}{P^{\alpha \beta}_{SD}  (\br,u = 0)}
\approx \frac{1}{1 + z_{\alpha \beta}(\br)}\;.
\eea
CPL stands for Coalescent Pair Locator, and in the following, we argue that this name may be appropriate as
the CPL can detect pairs involved in bonds, atomic shells, or lone pairs.
In the definition of  $z_{\alpha \beta}$ [see Eq.~(\ref{z})], we have set   $c=0.292$ 
in order to reproduce  the maximum value of the reference case in Fig.~\ref{H2}. Additional comments about this choice are given in Sec.~\ref{UEG}.
Adopting the nomenclature introduced by Perdew, we may regard the expression defined in Eq.~(\ref{CPL})
as a HGGA (i.e, Hyper-Generalized-Gradient-Approximation) because it makes use of the exact exchange energy density (i.e.,  the Slater potential).

Getting a ``maximum'' by means of the {\em approximate} 
CPL (see the dashed-red line in Fig.~\ref{H2}) is a non-trivial result. 
To appreciate this point fully, let us also consider 
the dot-dashed blue line in Fig.~\ref{H2}. This is another approximate CPL, 
which  is computed by  specializing the previous HGGA expression to the homogeneous gas  
and then using the resulting expression locally on the given inhomogeneous state.
This ``simplification'' is reminiscent of a Local Density Approximation (LDA) (see Sec.~\ref{UEG} for additional explanations).
Thus, we refer to this case as a CPL@LDA (regardless of which approximation has been used to generate the particle density itself).
The main point here is that: the desired maximum is then replaced by a {\em minimum}!
This is a consequence of the fact that  an LDA relates (too) directly to the local behavior of the particle density.

\begin{figure}[t]
\begin{center}
\includegraphics[width=0.9 \linewidth,angle=0]{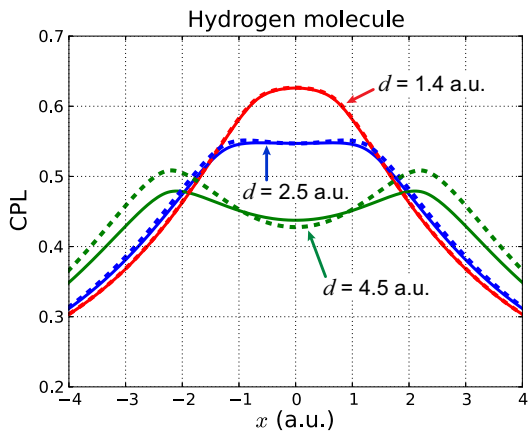}
\caption{\label{CPLfig}  The CPL  computed according to Eq.~(\ref{CPL}) with $c=0.292$ for H$_2$ at increasing atomic separation distance $d$.
The Slater potential, required as input, is obtained from  HF  (solid lines) and  full CI (dashed lines) densities.
Both the HF and CI calculations employ the Dunning gaussian basis set aug-cc-pv5Z.}\label{diss}
\end{center}
\vskip -0.5cm
\end{figure}

Now, let us follow the behavior of the CPL as the H$_2$ bond is stretched
in Fig.~\ref{diss}. In these calculations, we have used both HF and CI densities to visualize the dependence of the result on the quality of the input.
By increasing the interatomic distance, the local {\em maximum} at $x=0$ is eventually replaced by  a local {\em minimum} and two local maxima appear
around each H atom. Thus,  Fig.~\ref{diss}  suggests that by stretching the  molecule, the system splits into atoms. 
In the next section, we will
show that the CPL has maxima at the center of {\em isolated} atoms.

\section{Further Analyses}\label{Sec3}

In this section, we illustrate other salient differences of the CPL with respect to the ELF
by considering prototypical systems such as isolated atoms, small organic molecules, and the uniform electron gas. 

\subsection{Atomic Shells}\label{ArAtom}
\begin{figure}[h]
\begin{center}\label{Atom}
\includegraphics[width=1.05\columnwidth]{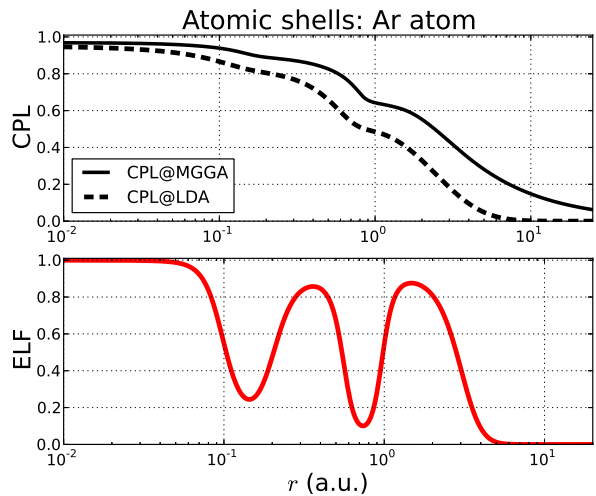}
\caption{The plots show the CPL (upper panel) and the ELF (lower panel) for an Argon atom.
CPL@MGGA (solid red line) was computed by adopting the BR89 approximation to the Slater potential \cite{BR89} in Eq.~(\ref{CPL}).
CPL@LDA (dot-dashed blue line) is computed according to Eq. (\ref{uCPL}).
KS@LDA results  obtained by means of the APE code \cite{APE} are used as an input to all the quantities.}\label{atom}
\end{center}
\vskip -0.5cm
\end{figure}

Electrons in atoms, we may say, are ``packed'' in shells. Figure~\ref{atom} shows how the CPL highlights these structures in the case of the Ar atom. 
Near the center of the atom, electrons get squeezed in the
1s orbital.
There the influence of the electron-electron interaction is on the average weaker relative to the nuclear attraction at the core and 
$\mathrm{CPL} \approx 1$. 
Moving away from the core, the CPL exhibits a wavy behavior. This renders the idea that electrons gain some space to avoid each other but in correspondence to the
shells, they get together once again. It is remarkable that, although
the particle density decreases monotonically and smoothly, the CPL shows some traces of the shells  even at the LDA level.
In systems with only one center such as atoms, the Slater potential can very effectively be estimated within the BR89 approximation \cite{BR89}.
Correspondingly, the calculation of the CPL  step down from the HGGA to the Meta-GGA (MGGA) rung.

Figure~\ref{atom} also shows the ELF which, at a first look, seems to be the champion in this case: it detects the atomic shells rather sharply.
But, we also notice that, an outer shells can have same (or higher) localization than an inner shell.
This somewhat awkward behavior in the localization of shell is not suggested by the CPL.

\subsection{Carbon-Carbon bonds}
\begin{figure}[h]
\includegraphics[width=1.05\columnwidth]{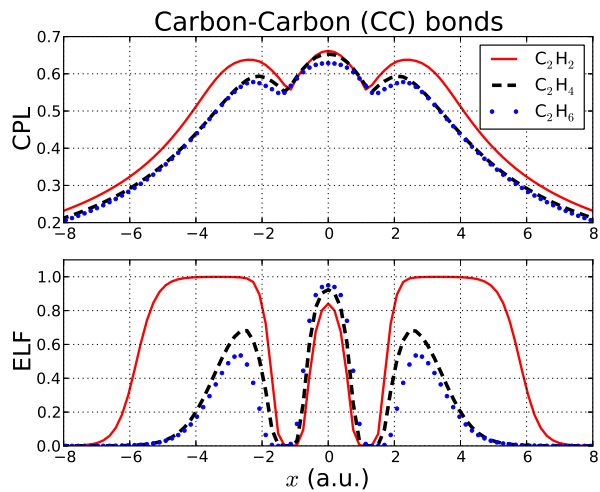}
\caption{{    The CPL  (upper panel) and ELF (lower panel)} for ethane, etylene and acetylene: having different CC bonds at equilibrium distances. 
 In all the cases, the CC bond is oriented along the $x$ direction. Input quantities are determined 
 from KS@LDA results  obtained by means of the Octopus code \cite{octopus} in a uniform spatial grid and pseudo potentials.}\label{CPL_mol_x} 
\vskip -0.5cm
\end{figure}

Let us consider again molecules: ethane, ethylene, and acetylene have single, double, and triple Carbon-Carbon (CC) bonds, respectively. The upper panel in Fig.~\ref{CPL_mol_x} 
shows the {    CPL}; the $x$ axis indicates the CC bond direction.
The calculations for these cases have been done using pseudo potentials, therefore the structures due to the internal electrons are ``washed'' out. Here, we are mainly interested in the structures due to the valence electrons (which are properly captured by the pseudo wavefunctions beyond a cutoff). 
The humps in the center highlight the presence of the bonds: the localization increases at larger bond coordination numbers. 
The two maxima 
on the side are due to the electrons of the Hydrogen atoms. For C$_2$H$_2$ the Hydrogen's centers are placed along the CC axis and, therefore, the electron
localization
results enhanced with respect to the other cases for which the Hydrogen's centers are displaced from the CC direction.

The lower panel of Fig.~\ref{CPL_mol_x} shows the ELF for the same molecules as in the upper panel.
It is apparent that the ELF, as for atoms, tends to give sharper features. But, unattractively, it assigns  higher localization to
bonds with lower coordination number. The CPL, however, 
does not yield surface plots 
with different topologies for bonds with different coordination numbers as in the case for the ELF. 

How about  $\pi-$conjugate systems? 
For benzene, in  plots  not shown here, the CPL does not seem to give any alternative information with respect to the
one already available with the ELF. Whether or not the ELF and the pair density itself can properly characterized the concept of aromaticity, we should point out, have been debated in the literature \cite{SteinmannPCCP11,BaerendsJCP10_b}.

\subsection{Polar bonds and lone pairs}

\begin{figure}[h]
\includegraphics[width=1.05\columnwidth]{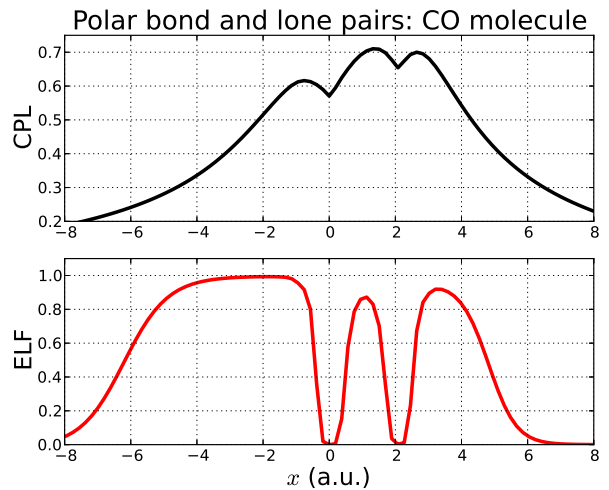}
\caption{{    The CPL  (upper panel) and ELF (lower panel) for the CO molecule.  
The molecule is oriented along the $x$ direction. The C atom is located at the origin and the O atom at $x=2.136$ a.u.
 Input quantities are determined from KS@LDA results by means of the Octopus code \cite{octopus} in a uniform spatial grid and with pseudo-potentials.} }\label{fig4b}
\vskip -0.5cm
\end{figure}

The molecule of carbon monoxide (CO) allows us to evaluate the performance of the CPL for both a polar bonds and two lone pairs.
As shown in Fig.~\ref{fig4b}, both the CPL and ELF pictures are overall satisfactory. The CPL, however, seems to  remark more distinctively  
the  higher electronegativity of  the oxygen atom with
respect to that of the carbon atoms. Note, moving  from C to O along the interatomic axis, the CPL  increases progressively while a reversed trend can be observed in the ELF. 
Furthermore, the central peak in the CPL is (slightly) more shifted toward the Oxygen atom than the peak in the ELF .

\newpage

\subsection{Uniform gas}\label{UEG}

First, let us recall that the reference to the uniform gas was introduced empirically in the definition of the ELF [see Eq.~(\ref{chi})]
in order to get meaningful plots. Moreover, $\mathrm{ELF} = 0.5$ for {\em any} uniform $n$. 
The concept of localization for a uniform gas is properly related to Wigner crystallization. This occurs at the Wigner-Seitz radius $r_s \sim 106$ \cite{WC1,WC2}, 
which corresponds to 
a correlated regime for which neither the ELF nor the CPL  -- by  construction -- can say much. Thus, in our analysis, we shall restrict to $ r_s \le 10$ to evaluate
further aspect of the CPL not fully discussed above.

For the uniform gas, a parametrization of the  pair-correlation function is available \cite{GSB}  -- here denoted as $g^{\alpha \beta}_{c,{\rm GSB}}(\br,u)$ --
to estimate the ratio $R$ [see Eq.~(\ref{R})] as follows
\ben\label{uR}
\mathrm{R}^{\rm unif} = 1 + g^{\alpha \beta}_{c,{\rm GSB}}(u=0)\;.
\een
This parametrization exploits Monte Carlo results and is expected to be accurate for densities corresponding to $r_s \le 10$. 
The CPL  [see Eq.~(\ref{CPL})] can be  evaluated with plain waves straightforwardly. As a result, we get
\ben\label{uCPL}
\mathrm{CPL}^{\rm unif}  = \frac{1}{1+   \frac{c}{3} (\frac{16 \pi}{3})^{\frac{2}{3}} r_s  }\;.
\een

\begin{figure}[h]
\includegraphics[width=0.9\columnwidth]{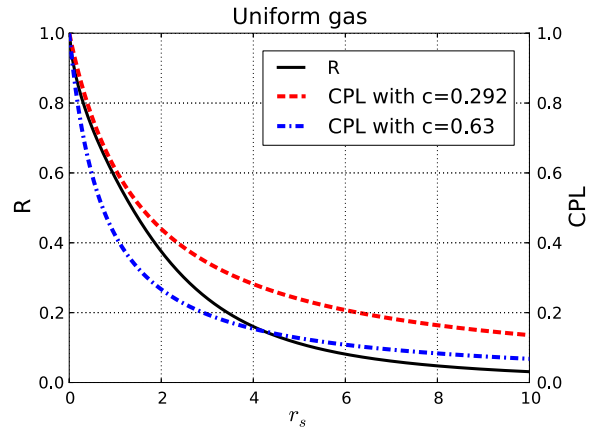}
\caption{The solid black curve provides us with an accurate reference for $R^{\rm unif}$ [Eq.~(\ref{uR})] within Wigner-Sitez radius $r_s \le 10$. 
The other curves show  the CPL for the uniform gas evaluated for different values of the constants $c$ [see Eq.~(\ref{uCPL})].}
\label{fUEG}
\vskip -0.5cm
\end{figure}

$\mathrm{R}^{\rm unif}$ and $\mathrm{CPL}^{\rm unif}$ are plotted in Fig.~\ref{fUEG}.
Fitting the height of the hump for H$_2$ in Fig.~\ref{H2}, we determined 
$c = 0.292$ (and this value was then used throughout in this work). We recall that  $c = 0.630$ in Becke's work 
reproduces the DFT correlation energies for the Helium atom \cite{cB88}.
Here, we see hat
$c = 0.292$ better reproduces the initial slope of $\mathrm{R}^{\rm unif}$ while  $c = 0.630$  improves  the agreement at higher $r_s$.
These results suggest us that: (i) for the considered range of densities, dynamical correlations -- as defined by the adopted model -- contribute to ${R}^{\rm unif}$ importantly; and 
(ii) treating $c$ as a constant is a drastic simplification, even in the case of a uniform gas. 
In general, $c$ should be replaced by  a density functional --
the search of which is postponed to future works.

In the high-density limit (which is the weakly correlated limit) $r_s  \rightarrow 0$ and, correctly, $\mathrm{CPL} \rightarrow 1$.  
In the opposite limit, $r_s  \rightarrow +\infty$ and $\mathrm{CPL} \rightarrow 0$.
Although tempting at first, however, $\mathrm{CPL} \sim 0$ cannot be taken as an indication of {\em locally} strong interactions as
this value is attained in the asymptotic region of any finite system.

\section{Conclusions and outlooks}\label{Sec4}

Bonds, shells, and lone pairs are important inhomogeneity features of atomistic systems. 
The interest in the evaluations of 
 these structures
 links usefully to the development of improved density functional approximations. 
In this work, we have proposed to probe them by exploiting the behavior of (a model of)
spatially-resolved dynamical correlations.
In particular, in favor of practicality, the {\em interacting} quantities were computed in a post-KS fashion.
We have focused on closed-shell systems,
for which static correlations are not expected to play a major role at equilibrium distances.
For the considered cases, we have seen that, the approach: (i) provides a spatially resolved description of the bond in H$_2$; 
(ii) points to a lower localization of the outer atomic shells;  (iii) suggests higher localization of multiple Carbon-Carbon bonds; and (iv)
distinctively remarks the location of elements with higher electronegativity  -- 
these are all points which would be missed in a description based solely on the ELF. 

Our procedure, admittedly, rests on simple ideas and models.
To enhance both transferability and accuracy, the parameter of the employed model (the constant  ``$c$'')  should be replaced with a proper density functional. 
Alternative separations of static and dynamical correlations (not inspired by DFT approximations) should also be explored.
Yet, we believe that, our analysis can usefully  suggests novel mixing  for {\em local} hybrids schemes or new
partition-DFT schemes.  Connections to recent Pair-DFT developments \cite{Gagliardi16} may also be particularly fruitful. 
Works along these lines are in progress.

\emph{Acknowledgements} --
The authors are grateful to Prof. Kieron Burke for making available data from Fig. 1 of Ref. \cite{BPE98}. 
SP acknowledges support by the European Community through the FP7's MC-IIF MODENADYNA, grant agreement No. 623413. DV acknowledges partial support from 
EU Centre  of Excellence MaX-Materials at the eXascale 
(H2020-EINFRA 2015, grant No. 676598).

\bibliography{paper_arxiv_res}

\begin{thebibliography}{40}%
\makeatletter
\providecommand \@ifxundefined [1]{%
 \@ifx{#1\undefined}
}%
\providecommand \@ifnum [1]{%
 \ifnum #1\expandafter \@firstoftwo
 \else \expandafter \@secondoftwo
 \fi
}%
\providecommand \@ifx [1]{%
 \ifx #1\expandafter \@firstoftwo
 \else \expandafter \@secondoftwo
 \fi
}%
\providecommand \natexlab [1]{#1}%
\providecommand \enquote  [1]{``#1''}%
\providecommand \bibnamefont  [1]{#1}%
\providecommand \bibfnamefont [1]{#1}%
\providecommand \citenamefont [1]{#1}%
\providecommand \href@noop [0]{\@secondoftwo}%
\providecommand \href [0]{\begingroup \@sanitize@url \@href}%
\providecommand \@href[1]{\@@startlink{#1}\@@href}%
\providecommand \@@href[1]{\endgroup#1\@@endlink}%
\providecommand \@sanitize@url [0]{\catcode `\\12\catcode `\$12\catcode
  `\&12\catcode `\#12\catcode `\^12\catcode `\_12\catcode `\%12\relax}%
\providecommand \@@startlink[1]{}%
\providecommand \@@endlink[0]{}%
\providecommand \url  [0]{\begingroup\@sanitize@url \@url }%
\providecommand \@url [1]{\endgroup\@href {#1}{\urlprefix }}%
\providecommand \urlprefix  [0]{URL }%
\providecommand \Eprint [0]{\href }%
\providecommand \doibase [0]{http://dx.doi.org/}%
\providecommand \selectlanguage [0]{\@gobble}%
\providecommand \bibinfo  [0]{\@secondoftwo}%
\providecommand \bibfield  [0]{\@secondoftwo}%
\providecommand \translation [1]{[#1]}%
\providecommand \BibitemOpen [0]{}%
\providecommand \bibitemStop [0]{}%
\providecommand \bibitemNoStop [0]{.\EOS\space}%
\providecommand \EOS [0]{\spacefactor3000\relax}%
\providecommand \BibitemShut  [1]{\csname bibitem#1\endcsname}%
\let\auto@bib@innerbib\@empty
\bibitem [{\citenamefont {Lewis}(1916)}]{Lewis}%
  \BibitemOpen
  \bibfield  {author} {\bibinfo {author} {\bibfnamefont {G.~N.}\ \bibnamefont
  {Lewis}},\ }\href {\doibase 10.1021/ja02261a002} {\bibfield  {journal}
  {\bibinfo  {journal} {Journal of the American Chemical Society}\ }\textbf
  {\bibinfo {volume} {38}},\ \bibinfo {pages} {762} (\bibinfo {year} {1916})},\
  \Eprint {http://arxiv.org/abs/http://dx.doi.org/10.1021/ja02261a002}
  {http://dx.doi.org/10.1021/ja02261a002} \BibitemShut {NoStop}%
\bibitem [{\citenamefont {Coulson}(1955)}]{Coulson55}%
  \BibitemOpen
  \bibfield  {author} {\bibinfo {author} {\bibfnamefont {C.~A.}\ \bibnamefont
  {Coulson}},\ }\href {\doibase 10.1039/JR9550002069} {\bibfield  {journal}
  {\bibinfo  {journal} {J. Chem. Soc.}\ ,\ \bibinfo {pages} {2069}} (\bibinfo
  {year} {1955})}\BibitemShut {NoStop}%
\bibitem [{\citenamefont {Bader}(1994)}]{Bader94}%
  \BibitemOpen
  \bibfield  {author} {\bibinfo {author} {\bibfnamefont {R.~F.~W.}\
  \bibnamefont {Bader}},\ }\href@noop {} {\emph {\bibinfo {title} {{Atoms in
  Molecules: A Quantum Theory (International Series of Monographs on
  Chemistry)}}}},\ \bibinfo {edition} {1st}\ ed.\ (\bibinfo  {publisher}
  {Clarendon Press},\ \bibinfo {year} {1994})\BibitemShut {NoStop}%
\bibitem [{\citenamefont {Bader}\ and\ \citenamefont
  {Stephens}(1975)}]{BaderXhole1}%
  \BibitemOpen
  \bibfield  {author} {\bibinfo {author} {\bibfnamefont {R.~F.~W.}\
  \bibnamefont {Bader}}\ and\ \bibinfo {author} {\bibfnamefont {M.~E.}\
  \bibnamefont {Stephens}},\ }\href {\doibase 10.1021/ja00859a001} {\bibfield
  {journal} {\bibinfo  {journal} {Journal of the American Chemical Society}\
  }\textbf {\bibinfo {volume} {97}},\ \bibinfo {pages} {7391} (\bibinfo {year}
  {1975})},\ \Eprint
  {http://arxiv.org/abs/http://dx.doi.org/10.1021/ja00859a001}
  {http://dx.doi.org/10.1021/ja00859a001} \BibitemShut {NoStop}%
\bibitem [{\citenamefont {Bader}\ \emph {et~al.}(1988)\citenamefont {Bader},
  \citenamefont {Gillespie},\ and\ \citenamefont {MacDougall}}]{BaderXhole2}%
  \BibitemOpen
  \bibfield  {author} {\bibinfo {author} {\bibfnamefont {R.~F.~W.}\
  \bibnamefont {Bader}}, \bibinfo {author} {\bibfnamefont {R.~J.}\ \bibnamefont
  {Gillespie}}, \ and\ \bibinfo {author} {\bibfnamefont {P.~J.}\ \bibnamefont
  {MacDougall}},\ }\href {\doibase 10.1021/ja00230a009} {\bibfield  {journal}
  {\bibinfo  {journal} {Journal of the American Chemical Society}\ }\textbf
  {\bibinfo {volume} {110}},\ \bibinfo {pages} {7329} (\bibinfo {year}
  {1988})},\ \Eprint
  {http://arxiv.org/abs/http://dx.doi.org/10.1021/ja00230a009}
  {http://dx.doi.org/10.1021/ja00230a009} \BibitemShut {NoStop}%
\bibitem [{\citenamefont {Becke}\ and\ \citenamefont {Edgecombe}(1990)}]{BE90}%
  \BibitemOpen
  \bibfield  {author} {\bibinfo {author} {\bibfnamefont {A.~D.}\ \bibnamefont
  {Becke}}\ and\ \bibinfo {author} {\bibfnamefont {K.~E.}\ \bibnamefont
  {Edgecombe}},\ }\href {\doibase 10.1063/1.458517} {\bibfield  {journal}
  {\bibinfo  {journal} {The Journal of Chemical Physics}\ }\textbf {\bibinfo
  {volume} {92}},\ \bibinfo {pages} {5397} (\bibinfo {year} {1990})},\ \Eprint
  {http://arxiv.org/abs/http://dx.doi.org/10.1063/1.458517}
  {http://dx.doi.org/10.1063/1.458517} \BibitemShut {NoStop}%
\bibitem [{\citenamefont {Savin}\ \emph {et~al.}(1992)\citenamefont {Savin},
  \citenamefont {Jepsen}, \citenamefont {Flad}, \citenamefont {Andersen},
  \citenamefont {Preuss},\ and\ \citenamefont {von Schnering}}]{Savin92}%
  \BibitemOpen
  \bibfield  {author} {\bibinfo {author} {\bibfnamefont {A.}~\bibnamefont
  {Savin}}, \bibinfo {author} {\bibfnamefont {O.}~\bibnamefont {Jepsen}},
  \bibinfo {author} {\bibfnamefont {J.}~\bibnamefont {Flad}}, \bibinfo {author}
  {\bibfnamefont {O.~K.}\ \bibnamefont {Andersen}}, \bibinfo {author}
  {\bibfnamefont {H.}~\bibnamefont {Preuss}}, \ and\ \bibinfo {author}
  {\bibfnamefont {H.~G.}\ \bibnamefont {von Schnering}},\ }\href {\doibase
  10.1002/anie.199201871} {\bibfield  {journal} {\bibinfo  {journal}
  {Angewandte Chemie International Edition in English}\ }\textbf {\bibinfo
  {volume} {31}},\ \bibinfo {pages} {187} (\bibinfo {year} {1992})}\BibitemShut
  {NoStop}%
\bibitem [{\citenamefont {{Silvi}}\ and\ \citenamefont
  {{Savin}}(1994)}]{Savin94}%
  \BibitemOpen
  \bibfield  {author} {\bibinfo {author} {\bibfnamefont {B.}~\bibnamefont
  {{Silvi}}}\ and\ \bibinfo {author} {\bibfnamefont {A.}~\bibnamefont
  {{Savin}}},\ }\href {\doibase 10.1038/371683a0} {\bibfield  {journal}
  {\bibinfo  {journal} {Nature}\ }\textbf {\bibinfo {volume} {371}},\ \bibinfo
  {pages} {683} (\bibinfo {year} {1994})}\BibitemShut {NoStop}%
\bibitem [{\citenamefont {Burnus}\ \emph {et~al.}(2005)\citenamefont {Burnus},
  \citenamefont {Marques},\ and\ \citenamefont {Gross}}]{BMG}%
  \BibitemOpen
  \bibfield  {author} {\bibinfo {author} {\bibfnamefont {T.}~\bibnamefont
  {Burnus}}, \bibinfo {author} {\bibfnamefont {M.~A.~L.}\ \bibnamefont
  {Marques}}, \ and\ \bibinfo {author} {\bibfnamefont {E.~K.~U.}\ \bibnamefont
  {Gross}},\ }\href {\doibase 10.1103/PhysRevA.71.010501} {\bibfield  {journal}
  {\bibinfo  {journal} {Phys. Rev. A}\ }\textbf {\bibinfo {volume} {71}},\
  \bibinfo {pages} {010501} (\bibinfo {year} {2005})}\BibitemShut {NoStop}%
\bibitem [{\citenamefont {Hodgson}\ \emph {et~al.}(2014)\citenamefont
  {Hodgson}, \citenamefont {Ramsden}, \citenamefont {Durrant},\ and\
  \citenamefont {Godby}}]{RG1}%
  \BibitemOpen
  \bibfield  {author} {\bibinfo {author} {\bibfnamefont {M.~J.~P.}\
  \bibnamefont {Hodgson}}, \bibinfo {author} {\bibfnamefont {J.~D.}\
  \bibnamefont {Ramsden}}, \bibinfo {author} {\bibfnamefont {T.~R.}\
  \bibnamefont {Durrant}}, \ and\ \bibinfo {author} {\bibfnamefont {R.~W.}\
  \bibnamefont {Godby}},\ }\href {\doibase 10.1103/PhysRevB.90.241107}
  {\bibfield  {journal} {\bibinfo  {journal} {Phys. Rev. B}\ }\textbf {\bibinfo
  {volume} {90}},\ \bibinfo {pages} {241107} (\bibinfo {year}
  {2014})}\BibitemShut {NoStop}%
\bibitem [{\citenamefont {Pittalis}\ \emph {et~al.}(2015)\citenamefont
  {Pittalis}, \citenamefont {Troiani}, \citenamefont {Rozzi},\ and\
  \citenamefont {Vignale}}]{Ent}%
  \BibitemOpen
  \bibfield  {author} {\bibinfo {author} {\bibfnamefont {S.}~\bibnamefont
  {Pittalis}}, \bibinfo {author} {\bibfnamefont {F.}~\bibnamefont {Troiani}},
  \bibinfo {author} {\bibfnamefont {C.~A.}\ \bibnamefont {Rozzi}}, \ and\
  \bibinfo {author} {\bibfnamefont {G.}~\bibnamefont {Vignale}},\ }\href
  {\doibase 10.1103/PhysRevB.91.075109} {\bibfield  {journal} {\bibinfo
  {journal} {Phys. Rev. B}\ }\textbf {\bibinfo {volume} {91}},\ \bibinfo
  {pages} {075109} (\bibinfo {year} {2015})}\BibitemShut {NoStop}%
\bibitem [{\citenamefont {Tao}\ \emph {et~al.}(2008)\citenamefont {Tao},
  \citenamefont {Vignale},\ and\ \citenamefont {Tokatly}}]{Tao}%
  \BibitemOpen
  \bibfield  {author} {\bibinfo {author} {\bibfnamefont {J.}~\bibnamefont
  {Tao}}, \bibinfo {author} {\bibfnamefont {G.}~\bibnamefont {Vignale}}, \ and\
  \bibinfo {author} {\bibfnamefont {I.~V.}\ \bibnamefont {Tokatly}},\ }\href
  {\doibase 10.1103/PhysRevLett.100.206405} {\bibfield  {journal} {\bibinfo
  {journal} {Phys. Rev. Lett.}\ }\textbf {\bibinfo {volume} {100}},\ \bibinfo
  {pages} {206405} (\bibinfo {year} {2008})}\BibitemShut {NoStop}%
\bibitem [{\citenamefont {{Durrant}}\ \emph {et~al.}(2015)\citenamefont
  {{Durrant}}, \citenamefont {{Hodgson}}, \citenamefont {{Ramsden}},\ and\
  \citenamefont {{Godby}}}]{RG2}%
  \BibitemOpen
  \bibfield  {author} {\bibinfo {author} {\bibfnamefont {T.~R.}\ \bibnamefont
  {{Durrant}}}, \bibinfo {author} {\bibfnamefont {M.~J.~P.}\ \bibnamefont
  {{Hodgson}}}, \bibinfo {author} {\bibfnamefont {J.~D.}\ \bibnamefont
  {{Ramsden}}}, \ and\ \bibinfo {author} {\bibfnamefont {R.~W.}\ \bibnamefont
  {{Godby}}},\ }\href@noop {} {\bibfield  {journal} {\bibinfo  {journal} {ArXiv
  e-prints}\ } (\bibinfo {year} {2015})},\ \Eprint
  {http://arxiv.org/abs/1505.07687} {arXiv:1505.07687 [cond-mat.mes-hall]}
  \BibitemShut {NoStop}%
\bibitem [{\citenamefont {Savin}(2005{\natexlab{a}})}]{SavinTheo05}%
  \BibitemOpen
  \bibfield  {author} {\bibinfo {author} {\bibfnamefont {A.}~\bibnamefont
  {Savin}},\ }\href {\doibase https://doi.org/10.1016/j.theochem.2005.02.034}
  {\bibfield  {journal} {\bibinfo  {journal} {Journal of Molecular Structure:
  THEOCHEM}\ }\textbf {\bibinfo {volume} {727}},\ \bibinfo {pages} {127 }
  (\bibinfo {year} {2005}{\natexlab{a}})},\ \bibinfo {note} {dedicated to Ramon
  Carbo\'o-Dorca on the Occasion of his 65th Birthday}\BibitemShut {NoStop}%
\bibitem [{\citenamefont {Savin}(2005{\natexlab{b}})}]{SavinJCS05}%
  \BibitemOpen
  \bibfield  {author} {\bibinfo {author} {\bibfnamefont {A.}~\bibnamefont
  {Savin}},\ }\href {\doibase 10.1007/BF02708351} {\bibfield  {journal}
  {\bibinfo  {journal} {Journal of Chemical Sciences}\ }\textbf {\bibinfo
  {volume} {117}},\ \bibinfo {pages} {473} (\bibinfo {year}
  {2005}{\natexlab{b}})}\BibitemShut {NoStop}%
\bibitem [{\citenamefont {Kohout}\ \emph {et~al.}(2004)\citenamefont {Kohout},
  \citenamefont {Pernal}, \citenamefont {Wagner},\ and\ \citenamefont
  {Grin}}]{Kohout04}%
  \BibitemOpen
  \bibfield  {author} {\bibinfo {author} {\bibfnamefont {M.}~\bibnamefont
  {Kohout}}, \bibinfo {author} {\bibfnamefont {K.}~\bibnamefont {Pernal}},
  \bibinfo {author} {\bibfnamefont {F.~R.}\ \bibnamefont {Wagner}}, \ and\
  \bibinfo {author} {\bibfnamefont {Y.}~\bibnamefont {Grin}},\ }\href {\doibase
  10.1007/s00214-004-0615-y} {\bibfield  {journal} {\bibinfo  {journal}
  {Theoretical Chemistry Accounts}\ }\textbf {\bibinfo {volume} {112}},\
  \bibinfo {pages} {453} (\bibinfo {year} {2004})}\BibitemShut {NoStop}%
\bibitem [{\citenamefont {Kohout}\ \emph {et~al.}(2005)\citenamefont {Kohout},
  \citenamefont {Pernal}, \citenamefont {Wagner},\ and\ \citenamefont
  {Grin}}]{Kohout05}%
  \BibitemOpen
  \bibfield  {author} {\bibinfo {author} {\bibfnamefont {M.}~\bibnamefont
  {Kohout}}, \bibinfo {author} {\bibfnamefont {K.}~\bibnamefont {Pernal}},
  \bibinfo {author} {\bibfnamefont {F.~R.}\ \bibnamefont {Wagner}}, \ and\
  \bibinfo {author} {\bibfnamefont {Y.}~\bibnamefont {Grin}},\ }\href {\doibase
  10.1007/s00214-005-0671-y} {\bibfield  {journal} {\bibinfo  {journal}
  {Theoretical Chemistry Accounts}\ }\textbf {\bibinfo {volume} {113}},\
  \bibinfo {pages} {287} (\bibinfo {year} {2005})}\BibitemShut {NoStop}%
\bibitem [{\citenamefont {Feixas}\ \emph {et~al.}(2010)\citenamefont {Feixas},
  \citenamefont {Matito}, \citenamefont {Duran}, \citenamefont {Sol\`a},\ and\
  \citenamefont {Silvi}}]{Silvi2010}%
  \BibitemOpen
  \bibfield  {author} {\bibinfo {author} {\bibfnamefont {F.}~\bibnamefont
  {Feixas}}, \bibinfo {author} {\bibfnamefont {E.}~\bibnamefont {Matito}},
  \bibinfo {author} {\bibfnamefont {M.}~\bibnamefont {Duran}}, \bibinfo
  {author} {\bibfnamefont {M.}~\bibnamefont {Sol\`a}}, \ and\ \bibinfo {author}
  {\bibfnamefont {B.}~\bibnamefont {Silvi}},\ }\href {\doibase
  10.1021/ct1003548} {\bibfield  {journal} {\bibinfo  {journal} {Journal of
  Chemical Theory and Computation}\ }\textbf {\bibinfo {volume} {6}},\ \bibinfo
  {pages} {2736} (\bibinfo {year} {2010})},\ \bibinfo {note} {pMID: 26616075},\
  \Eprint {http://arxiv.org/abs/http://dx.doi.org/10.1021/ct1003548}
  {http://dx.doi.org/10.1021/ct1003548} \BibitemShut {NoStop}%
\bibitem [{\citenamefont {Helgaker}\ \emph {et~al.}(2000)\citenamefont
  {Helgaker}, \citenamefont {Jorgensen},\ and\ \citenamefont
  {Olsen}}]{helgaker2000}%
  \BibitemOpen
  \bibfield  {author} {\bibinfo {author} {\bibfnamefont {T.}~\bibnamefont
  {Helgaker}}, \bibinfo {author} {\bibfnamefont {P.}~\bibnamefont {Jorgensen}},
  \ and\ \bibinfo {author} {\bibfnamefont {J.}~\bibnamefont {Olsen}},\
  }\href@noop {} {\emph {\bibinfo {title} {Molecular Electronic-structure
  Theory}}}\ (\bibinfo  {publisher} {Wiley},\ \bibinfo {year}
  {2000})\BibitemShut {NoStop}%
\bibitem [{\citenamefont {Becke}(1988)}]{cB88}%
  \BibitemOpen
  \bibfield  {author} {\bibinfo {author} {\bibfnamefont {A.~D.}\ \bibnamefont
  {Becke}},\ }\href {\doibase 10.1063/1.454274} {\bibfield  {journal} {\bibinfo
   {journal} {The Journal of Chemical Physics}\ }\textbf {\bibinfo {volume}
  {88}},\ \bibinfo {pages} {1053} (\bibinfo {year} {1988})},\ \Eprint
  {http://arxiv.org/abs/http://dx.doi.org/10.1063/1.454274}
  {http://dx.doi.org/10.1063/1.454274} \BibitemShut {NoStop}%
\bibitem [{\citenamefont {Gallegos}\ \emph {et~al.}(2005)\citenamefont
  {Gallegos}, \citenamefont {Carb\'o-Dorca}, \citenamefont {Lodier},
  \citenamefont {Canc\`es},\ and\ \citenamefont {Savin}}]{SavinJCC05}%
  \BibitemOpen
  \bibfield  {author} {\bibinfo {author} {\bibfnamefont {A.}~\bibnamefont
  {Gallegos}}, \bibinfo {author} {\bibfnamefont {R.}~\bibnamefont
  {Carb\'o-Dorca}}, \bibinfo {author} {\bibfnamefont {F.}~\bibnamefont
  {Lodier}}, \bibinfo {author} {\bibfnamefont {E.}~\bibnamefont {Canc\`es}}, \
  and\ \bibinfo {author} {\bibfnamefont {A.}~\bibnamefont {Savin}},\ }\href
  {\doibase 10.1002/jcc.20180} {\bibfield  {journal} {\bibinfo  {journal}
  {Journal of Computational Chemistry}\ }\textbf {\bibinfo {volume} {26}},\
  \bibinfo {pages} {455} (\bibinfo {year} {2005})}\BibitemShut {NoStop}%
\bibitem [{\citenamefont {Men\'endez}\ \emph {et~al.}(2015)\citenamefont
  {Men\'endez}, \citenamefont {Pend\'as}, \citenamefont {Bra{\"i}da},\ and\
  \citenamefont {Savin}}]{SavinCTC15}%
  \BibitemOpen
  \bibfield  {author} {\bibinfo {author} {\bibfnamefont {M.}~\bibnamefont
  {Men\'endez}}, \bibinfo {author} {\bibfnamefont {A.~M.}\ \bibnamefont
  {Pend\'as}}, \bibinfo {author} {\bibfnamefont {B.}~\bibnamefont
  {Bra{\"i}da}}, \ and\ \bibinfo {author} {\bibfnamefont {A.}~\bibnamefont
  {Savin}},\ }\href {\doibase https://doi.org/10.1016/j.comptc.2014.10.004}
  {\bibfield  {journal} {\bibinfo  {journal} {Computational and Theoretical
  Chemistry}\ }\textbf {\bibinfo {volume} {1053}},\ \bibinfo {pages} {142 }
  (\bibinfo {year} {2015})},\ \bibinfo {note} {special Issue: Understanding
  structure and reactivity from topology and beyond}\BibitemShut {NoStop}%
\bibitem [{Note1()}]{Note1}%
  \BibitemOpen
  \bibinfo {note} {Our Kohn-Sham (or Hartree-Fock) calculations are intended as
  means to compute a ``descriptor'' of the electronic structure of the system
  under consideration and not necessarily to estimate the best possible total
  energies -- thus we avoid to spuriously break spin symmetry.}\BibitemShut
  {Stop}%
\bibitem [{Note2()}]{Note2}%
  \BibitemOpen
  \bibinfo {note} {Given a generic function $f({{\protect \bf r}_1,{\protect
  \bf r}_2})$, its spherical average at position ${\protect \bf r}$ for a pair
  at distance $u$ may be defined as the integral $f ({\protect \bf r},u) :=
  \protect \frac {1}{4 \pi } \DOTSI \intop \ilimits@ d \Omega _{{\protect \bf
  u}}~ f ( {\protect \bf r}, {\protect \bf r}+ {\protect \bf u})$ where $d
  \Omega _{{\protect \bf u}}$ is an element of the solid angle.}\BibitemShut
  {Stop}%
\bibitem [{Note3()}]{Note3}%
  \BibitemOpen
  \bibinfo {note} {The {\protect \em exact} KS Slater determinant in (Spin-)DFT
  reproduces the interacting particle (spin-)density exactly.}\BibitemShut
  {Stop}%
\bibitem [{\citenamefont {Savin}\ \emph {et~al.}(1997)\citenamefont {Savin},
  \citenamefont {Nesper}, \citenamefont {Wengert},\ and\ \citenamefont
  {F\"assler}}]{Savinreview}%
  \BibitemOpen
  \bibfield  {author} {\bibinfo {author} {\bibfnamefont {A.}~\bibnamefont
  {Savin}}, \bibinfo {author} {\bibfnamefont {R.}~\bibnamefont {Nesper}},
  \bibinfo {author} {\bibfnamefont {S.}~\bibnamefont {Wengert}}, \ and\
  \bibinfo {author} {\bibfnamefont {T.~F.}\ \bibnamefont {F\"assler}},\ }\href
  {\doibase 10.1002/anie.199718081} {\bibfield  {journal} {\bibinfo  {journal}
  {Angewandte Chemie International Edition in English}\ }\textbf {\bibinfo
  {volume} {36}},\ \bibinfo {pages} {1808} (\bibinfo {year}
  {1997})}\BibitemShut {NoStop}%
\bibitem [{\citenamefont {Burke}\ \emph {et~al.}(1998)\citenamefont {Burke},
  \citenamefont {Perdew},\ and\ \citenamefont {Ernzerhof}}]{BPE98}%
  \BibitemOpen
  \bibfield  {author} {\bibinfo {author} {\bibfnamefont {K.}~\bibnamefont
  {Burke}}, \bibinfo {author} {\bibfnamefont {J.~P.}\ \bibnamefont {Perdew}}, \
  and\ \bibinfo {author} {\bibfnamefont {M.}~\bibnamefont {Ernzerhof}},\ }\href
  {\doibase 10.1063/1.476976} {\bibfield  {journal} {\bibinfo  {journal} {The
  Journal of Chemical Physics}\ }\textbf {\bibinfo {volume} {109}},\ \bibinfo
  {pages} {3760} (\bibinfo {year} {1998})},\ \Eprint
  {http://arxiv.org/abs/http://dx.doi.org/10.1063/1.476976}
  {http://dx.doi.org/10.1063/1.476976} \BibitemShut {NoStop}%
\bibitem [{Note4()}]{Note4}%
  \BibitemOpen
  \bibinfo {note} {Yet, it has been pointed out in the literature that use of
  HF densities can produce counterintuitive features in the corresponding
  on-top hole \cite {BaerendsJCP10_a}.}\BibitemShut {Stop}%
\bibitem [{\citenamefont {Schuchardt}\ \emph {et~al.}(2007)\citenamefont
  {Schuchardt}, \citenamefont {Didier}, \citenamefont {Elsethagen},
  \citenamefont {Sun}, \citenamefont {Gurumoorthi}, \citenamefont {Chase},
  \citenamefont {Li},\ and\ \citenamefont {Windus}}]{schuchardt2007basis}%
  \BibitemOpen
  \bibfield  {author} {\bibinfo {author} {\bibfnamefont {K.~L.}\ \bibnamefont
  {Schuchardt}}, \bibinfo {author} {\bibfnamefont {B.~T.}\ \bibnamefont
  {Didier}}, \bibinfo {author} {\bibfnamefont {T.}~\bibnamefont {Elsethagen}},
  \bibinfo {author} {\bibfnamefont {L.}~\bibnamefont {Sun}}, \bibinfo {author}
  {\bibfnamefont {V.}~\bibnamefont {Gurumoorthi}}, \bibinfo {author}
  {\bibfnamefont {J.}~\bibnamefont {Chase}}, \bibinfo {author} {\bibfnamefont
  {J.}~\bibnamefont {Li}}, \ and\ \bibinfo {author} {\bibfnamefont {T.~L.}\
  \bibnamefont {Windus}},\ }\href@noop {} {\bibfield  {journal} {\bibinfo
  {journal} {Journal of chemical information and modeling}\ }\textbf {\bibinfo
  {volume} {47}},\ \bibinfo {pages} {1045} (\bibinfo {year}
  {2007})}\BibitemShut {NoStop}%
\bibitem [{\citenamefont {K\"ummel}\ and\ \citenamefont {Kronik}(2008)}]{OEP}%
  \BibitemOpen
  \bibfield  {author} {\bibinfo {author} {\bibfnamefont {S.}~\bibnamefont
  {K\"ummel}}\ and\ \bibinfo {author} {\bibfnamefont {L.}~\bibnamefont
  {Kronik}},\ }\href {\doibase 10.1103/RevModPhys.80.3} {\bibfield  {journal}
  {\bibinfo  {journal} {Rev. Mod. Phys.}\ }\textbf {\bibinfo {volume} {80}},\
  \bibinfo {pages} {3} (\bibinfo {year} {2008})}\BibitemShut {NoStop}%
\bibitem [{\citenamefont {Becke}\ and\ \citenamefont {Roussel}(1989)}]{BR89}%
  \BibitemOpen
  \bibfield  {author} {\bibinfo {author} {\bibfnamefont {A.~D.}\ \bibnamefont
  {Becke}}\ and\ \bibinfo {author} {\bibfnamefont {M.~R.}\ \bibnamefont
  {Roussel}},\ }\href {\doibase 10.1103/PhysRevA.39.3761} {\bibfield  {journal}
  {\bibinfo  {journal} {Phys. Rev. A}\ }\textbf {\bibinfo {volume} {39}},\
  \bibinfo {pages} {3761} (\bibinfo {year} {1989})}\BibitemShut {NoStop}%
\bibitem [{\citenamefont {Oliveira}\ and\ \citenamefont
  {Nogueira}(2008)}]{APE}%
  \BibitemOpen
  \bibfield  {author} {\bibinfo {author} {\bibfnamefont {M.~J.}\ \bibnamefont
  {Oliveira}}\ and\ \bibinfo {author} {\bibfnamefont {F.}~\bibnamefont
  {Nogueira}},\ }\href {\doibase http://doi.org/10.1016/j.cpc.2007.11.003}
  {\bibfield  {journal} {\bibinfo  {journal} {Computer Physics Communications}\
  }\textbf {\bibinfo {volume} {178}},\ \bibinfo {pages} {524 } (\bibinfo {year}
  {2008})}\BibitemShut {NoStop}%
\bibitem [{\citenamefont {Castro}\ \emph {et~al.}(2006)\citenamefont {Castro},
  \citenamefont {Appel}, \citenamefont {Oliveira}, \citenamefont {Rozzi},
  \citenamefont {Andrade}, \citenamefont {Lorenzen}, \citenamefont {Marques},
  \citenamefont {Gross},\ and\ \citenamefont {Rubio}}]{octopus}%
  \BibitemOpen
  \bibfield  {author} {\bibinfo {author} {\bibfnamefont {A.}~\bibnamefont
  {Castro}}, \bibinfo {author} {\bibfnamefont {H.}~\bibnamefont {Appel}},
  \bibinfo {author} {\bibfnamefont {M.}~\bibnamefont {Oliveira}}, \bibinfo
  {author} {\bibfnamefont {C.~A.}\ \bibnamefont {Rozzi}}, \bibinfo {author}
  {\bibfnamefont {X.}~\bibnamefont {Andrade}}, \bibinfo {author} {\bibfnamefont
  {F.}~\bibnamefont {Lorenzen}}, \bibinfo {author} {\bibfnamefont {M.~A.~L.}\
  \bibnamefont {Marques}}, \bibinfo {author} {\bibfnamefont {E.~K.~U.}\
  \bibnamefont {Gross}}, \ and\ \bibinfo {author} {\bibfnamefont
  {A.}~\bibnamefont {Rubio}},\ }\href {\doibase 10.1002/pssb.200642067}
  {\bibfield  {journal} {\bibinfo  {journal} {physica status solidi (b)}\
  }\textbf {\bibinfo {volume} {243}},\ \bibinfo {pages} {2465} (\bibinfo {year}
  {2006})}\BibitemShut {NoStop}%
\bibitem [{\citenamefont {Steinmann}\ \emph {et~al.}(2011)\citenamefont
  {Steinmann}, \citenamefont {Mo},\ and\ \citenamefont
  {Corminboeuf}}]{SteinmannPCCP11}%
  \BibitemOpen
  \bibfield  {author} {\bibinfo {author} {\bibfnamefont {S.~N.}\ \bibnamefont
  {Steinmann}}, \bibinfo {author} {\bibfnamefont {Y.}~\bibnamefont {Mo}}, \
  and\ \bibinfo {author} {\bibfnamefont {C.}~\bibnamefont {Corminboeuf}},\
  }\href {\doibase 10.1039/C1CP21055F} {\bibfield  {journal} {\bibinfo
  {journal} {Phys. Chem. Chem. Phys.}\ }\textbf {\bibinfo {volume} {13}},\
  \bibinfo {pages} {20584} (\bibinfo {year} {2011})}\BibitemShut {NoStop}%
\bibitem [{\citenamefont {Wang}\ and\ \citenamefont
  {Baerends}(2015)}]{BaerendsJCP10_b}%
  \BibitemOpen
  \bibfield  {author} {\bibinfo {author} {\bibfnamefont {J.}~\bibnamefont
  {Wang}}\ and\ \bibinfo {author} {\bibfnamefont {E.~J.}\ \bibnamefont
  {Baerends}},\ }\href {\doibase 10.1063/1.4921725} {\bibfield  {journal}
  {\bibinfo  {journal} {The Journal of Chemical Physics}\ }\textbf {\bibinfo
  {volume} {142}},\ \bibinfo {pages} {204311} (\bibinfo {year} {2015})},\
  \Eprint {http://arxiv.org/abs/https://doi.org/10.1063/1.4921725}
  {https://doi.org/10.1063/1.4921725} \BibitemShut {NoStop}%
\bibitem [{\citenamefont {Ceperley}\ and\ \citenamefont {Alder}(1980)}]{WC1}%
  \BibitemOpen
  \bibfield  {author} {\bibinfo {author} {\bibfnamefont {D.~M.}\ \bibnamefont
  {Ceperley}}\ and\ \bibinfo {author} {\bibfnamefont {B.~J.}\ \bibnamefont
  {Alder}},\ }\href {\doibase 10.1103/PhysRevLett.45.566} {\bibfield  {journal}
  {\bibinfo  {journal} {Phys. Rev. Lett.}\ }\textbf {\bibinfo {volume} {45}},\
  \bibinfo {pages} {566} (\bibinfo {year} {1980})}\BibitemShut {NoStop}%
\bibitem [{\citenamefont {Drummond}\ \emph {et~al.}(2004)\citenamefont
  {Drummond}, \citenamefont {Radnai}, \citenamefont {Trail}, \citenamefont
  {Towler},\ and\ \citenamefont {Needs}}]{WC2}%
  \BibitemOpen
  \bibfield  {author} {\bibinfo {author} {\bibfnamefont {N.~D.}\ \bibnamefont
  {Drummond}}, \bibinfo {author} {\bibfnamefont {Z.}~\bibnamefont {Radnai}},
  \bibinfo {author} {\bibfnamefont {J.~R.}\ \bibnamefont {Trail}}, \bibinfo
  {author} {\bibfnamefont {M.~D.}\ \bibnamefont {Towler}}, \ and\ \bibinfo
  {author} {\bibfnamefont {R.~J.}\ \bibnamefont {Needs}},\ }\href {\doibase
  10.1103/PhysRevB.69.085116} {\bibfield  {journal} {\bibinfo  {journal} {Phys.
  Rev. B}\ }\textbf {\bibinfo {volume} {69}},\ \bibinfo {pages} {085116}
  (\bibinfo {year} {2004})}\BibitemShut {NoStop}%
\bibitem [{\citenamefont {Gori-Giorgi}\ \emph {et~al.}(2000)\citenamefont
  {Gori-Giorgi}, \citenamefont {Sacchetti},\ and\ \citenamefont
  {Bachelet}}]{GSB}%
  \BibitemOpen
  \bibfield  {author} {\bibinfo {author} {\bibfnamefont {P.}~\bibnamefont
  {Gori-Giorgi}}, \bibinfo {author} {\bibfnamefont {F.}~\bibnamefont
  {Sacchetti}}, \ and\ \bibinfo {author} {\bibfnamefont {G.~B.}\ \bibnamefont
  {Bachelet}},\ }\href {\doibase 10.1103/PhysRevB.61.7353} {\bibfield
  {journal} {\bibinfo  {journal} {Phys. Rev. B}\ }\textbf {\bibinfo {volume}
  {61}},\ \bibinfo {pages} {7353} (\bibinfo {year} {2000})}\BibitemShut
  {NoStop}%
\bibitem [{\citenamefont {Gagliardi}\ \emph {et~al.}(2017)\citenamefont
  {Gagliardi}, \citenamefont {Truhlar}, \citenamefont {Li~Manni}, \citenamefont
  {Carlson}, \citenamefont {Hoyer},\ and\ \citenamefont {Bao}}]{Gagliardi16}%
  \BibitemOpen
  \bibfield  {author} {\bibinfo {author} {\bibfnamefont {L.}~\bibnamefont
  {Gagliardi}}, \bibinfo {author} {\bibfnamefont {D.~G.}\ \bibnamefont
  {Truhlar}}, \bibinfo {author} {\bibfnamefont {G.}~\bibnamefont {Li~Manni}},
  \bibinfo {author} {\bibfnamefont {R.~K.}\ \bibnamefont {Carlson}}, \bibinfo
  {author} {\bibfnamefont {C.~E.}\ \bibnamefont {Hoyer}}, \ and\ \bibinfo
  {author} {\bibfnamefont {J.~L.}\ \bibnamefont {Bao}},\ }\href {\doibase
  10.1021/acs.accounts.6b00471} {\bibfield  {journal} {\bibinfo  {journal}
  {Accounts of Chemical Research}\ }\textbf {\bibinfo {volume} {50}},\ \bibinfo
  {pages} {66} (\bibinfo {year} {2017})},\ \bibinfo {note} {pMID: 28001359},\
  \Eprint {http://arxiv.org/abs/https://doi.org/10.1021/acs.accounts.6b00471}
  {https://doi.org/10.1021/acs.accounts.6b00471} \BibitemShut {NoStop}%
\bibitem [{\citenamefont {Wang}\ \emph {et~al.}(2010)\citenamefont {Wang},
  \citenamefont {Kim},\ and\ \citenamefont {Baerends}}]{BaerendsJCP10_a}%
  \BibitemOpen
  \bibfield  {author} {\bibinfo {author} {\bibfnamefont {J.}~\bibnamefont
  {Wang}}, \bibinfo {author} {\bibfnamefont {K.~S.}\ \bibnamefont {Kim}}, \
  and\ \bibinfo {author} {\bibfnamefont {E.~J.}\ \bibnamefont {Baerends}},\
  }\href {\doibase 10.1063/1.3429608} {\bibfield  {journal} {\bibinfo
  {journal} {The Journal of Chemical Physics}\ }\textbf {\bibinfo {volume}
  {132}},\ \bibinfo {pages} {204102} (\bibinfo {year} {2010})},\ \Eprint
  {http://arxiv.org/abs/https://doi.org/10.1063/1.3429608}
  {https://doi.org/10.1063/1.3429608} \BibitemShut {NoStop}%
\end{thebibliography}%

\end{document}